\documentclass[aps,prl,twocolumn,groupedaddress,floats,showpacs]{revtex4}

\usepackage{amsmath}
\usepackage[dvips]{graphics}
\usepackage{hyperref}
\usepackage{epsf}

\newcommand{\e}{\varepsilon}

\renewcommand{\vec}[1]{\mathbf{#1}}
\renewcommand{\vr}{\vec{r}}
\newcommand{\vm}{\vec{m}}
\newcommand{\ve}{\vec{e}}
\newcommand{\vsigma}{\mbox{\boldmath $\sigma$}}
\newcommand{\vA}{\vec{A}}
\newcommand{\vnabla}{\mbox{\boldmath $\nabla$}}

\begin{document}

\title{Enhanced triplet Andreev reflection off a domain wall in
  a lateral geometry} 
\author{Joern N.\ Kupferschmidt, Piet W.\ Brouwer}
 \affiliation{Laboratory of Atomic and Solid State Physics, Cornell
  University, Ithaca, NY 14853-2501, USA}
\date{April 24, 2009}

\begin{abstract}

We find that the triplet Andreev reflection amplitude at the interface
between a half-metal and an $s$-wave superconductor in the presence of
a domain wall is significantly enhanced if the half metal is a thin
film, rather than an extended magnet. The enhancement is by a factor
$l_{\rm d}/d$, where $l_{\rm d}$ is the width of the domain wall and
$d$ the film thickness. We conclude that in a lateral geometry,
domain walls can be an effective source of the triplet proximity
effect.


\end{abstract}

\pacs{74.45.+c,74.50.+r,74.78.Na,75.70.Cn}

\maketitle

A normal metal inherits superconducting properties if it is in
electrical contact to a superconductor. This ``superconductor
proximity effect'' is mediated by Andreev reflection
\cite{andreev1964}, the process in
which an electron incident from the normal metal is reflected as a
hole at the normal-metal--superconductor interface. 
At
the Fermi energy, phase coherence between the electron and the Andreev 
reflected hole is preserved over long distances, which is the reason 
why the induced superconducting correlations exist deep 
inside the normal metal.

At the interface between a ferromagnet and a superconductor, majority
electrons (electrons with their spin parallel to the magnetization
direction $\vm$) are Andreev reflected as minority holes and vice
versa \cite{endnote24}.
Since phases of majority electrons and minority holes are
not correlated, the proximity effect becomes effectively short-range 
in a ferromagnet. The situation is even more extreme in a half 
metal, a material in which only majority charge carriers exist. At 
a half-metal--superconductor interface Andreev reflection of majority
electrons is strongly suppressed, simply because of the absence of 
minority holes.

It was realized by Bergeret {\em et al.} \cite{bergeret2001}
(see also Ref.\ \cite{kadigrobov2001})
that the situation is entirely different if spin-rotation 
symmetry around the (mean) magnetization direction at the superconductor
interface is broken: In that case, majority electrons
may be reflected as majority holes. 
Since there is phase coherence
between majority electrons and majority holes, the resulting ``triplet
proximity effect'' can penetrate ferromagnets or half metals about the
same distance as the standard proximity effect penetrates normal
metals \cite{bergeret2005}. 
Various experiments have hinted at the existence of this 
effect \cite{sosnin2006,keizer2006,krivoruchko2007,yates2007}, the most striking of 
which is the observation of a Josephson current through a $\mu$m long 
link of the half metal $\mbox{CrO}_2$ by Keizer {\em et al.}
\cite{keizer2006}. 

There have been various proposals for the origin of the broken
spin-rotation symmetry needed for the existence of the triplet 
proximity effect. One
possibility is an artificial structure, in which there is a thin
ferromagnetic or half-metallic spacer layer at the interface with a
magnetization direction different from that of the bulk magnet 
\cite{asano2007,beri2009}. 
A second possibility is a magnetically disordered or ``spin-active''
interface
\cite{eschrig2003,eschrig2008}. 
Finally, the triplet proximity effect can
be caused by variations of the magnetization direction $\vm$
associated with a domain wall, either perpendicular \cite{volkov2008}
or parallel to the superconductor interface \cite{volkov2005}.

\begin{figure}
\epsfxsize=0.99\hsize
\epsffile{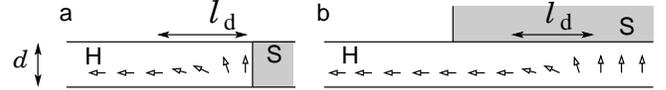}
\caption{\label{fig:1} Serial (a) and lateral (b)
superconductor--half-metal junction with a domain wall. The Andreev
reflection amplitude in (b) is enhanced with respect to that of (a) by
a large factor $\sim l_{\rm d}/d$.}
\end{figure}

Although domain walls occur generically in half metals and ferromagnets,
at first sight they are an unlikely source of the triplet
proximity effect, because (1) only domain walls that happen to be
immediately at the superconductor interface can contribute
to the triplet proximity effect and (2) the spin-flip scattering
amplitude in a domain wall is small as $1/(k_{\rm F} l_{\rm d})$,
where $k_{\rm F}$ is the Fermi wavelength and $l_{\rm d}$ the width of
the domain wall \cite{bergeret2005}. 
%
This severely restricts the magnitude and range of the triplet
proximity effect mediated by domain walls in the geometry of Fig.\
\ref{fig:1}a, in which
the half metal or ferromagnet and the superconductor are placed ``in
series''.

In this letter, we show that both limitations are
absent in a different geometry, shown in
Fig.\ \ref{fig:1}b, in which the superconductor is laterally coupled
to a magnetic film with a domain wall for which the direction of the
magnetization variation is parallel to the interface. 
Although the lateral geometry has received as good as no
theoretical attention  --- most theoretical works deal with the
serial geometry of Fig.\ \ref{fig:1}a ---, it is the relevant geometry for
the experiment of Ref.\ \cite{keizer2006}. For the lateral
geometry we find that 
majority electrons have an amplitude $r_{\rm he}$ for
Andreev reflection as majority holes that scales proportional to 
$(k_{\rm F} d)^{-1}$, where $d$ is the thickness of the magnetic
film, independent of the domain wall width $l_{\rm d}$ and independent 
of the precise location or orientation of the domain wall. 

We focus the discussion of this letter on the case of a half-metallic 
film. This is not only most relevant for the experiment of Ref.\ 
\cite{keizer2006}, it also allows for an easy identification of the 
triplet proximity effect because in a half metal the mere
existence of Andreev reflection is already a signature of the triplet 
proximity effect \cite{eschrig2003}. 
(In a ferromagnet, Andreev reflection also takes place in the absence
of the triplet proximity effect.) Our work complements previous
studies of the triplet proximity effect in the presence of
domain walls in ferromagnets in the limit of weak exchange fields
\cite{volkov2005}.

In the remainder of this letter, we present a calculation of the
Andreev reflection amplitude $r_{\rm he}$ for the
lateral geometry. We also discuss two applications: The two-terminal
conductance between the half metal and the superconductor in the
lateral geometry, and the Josephson effect in a 
superconductor--half-metal--superconductor junction.

\begin{figure}
\epsfxsize=0.8\hsize
\epsffile{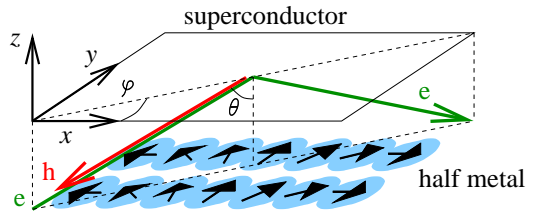}
\caption{\label{fig:2} (Color online) Half-metal--superconductor
  interface with a domain wall. An electron (e) incident on the
  interface is either normally reflected, or Andreev reflected as a
  hole (h). The Andreev reflection amplitude $r_{\rm he}$ for this situation
  given by Eq.\ (\ref{eq:rhe}) of the main text.}
\end{figure}

\textit{Amplitude for a single Andreev reflection.} We first calculate
the amplitude $r_{\rm he}$ for a single Andreev reflection off a
domain wall perpendicular to the half-metal--superconductor
interface. Quasiparticle excitations near the interface are described
by the Bogoliubov-de Gennes equation
\begin{equation} 
\left( \begin{array}{cc} 
  \hat H   &   i \Delta e^{i \phi} \sigma_2 \\
- i \Delta e^{-i \phi} \sigma_2 & - \hat H^*  
\end{array} \right) \Psi = \e \Psi,
\end{equation}
where $\Psi$ is a four-component wavefunction with components for the
electron/hole and spin degrees of freedom and 
and $\Delta e^{i \phi}$ is the superconducting order
parameter. We choose coordinates such that the
half-metal--superconductor interface is the plane $z=0$ and the
magnetization direction $\vm$ in the half metal varies in the $x$
direction, see Fig.\ \ref{fig:2}.
In the superconductor ($z > 0$), we take the Hamiltonian $\hat H$ to be
$\hat H = p^2/2 m_{\rm S} - \varepsilon_{\rm F,S}$, where $m_{\rm S}$
and $\varepsilon_{\rm F,S} = \hbar^2 k_{\rm S}^2/2 m_{\rm S}$ 
are the effective mass and Fermi energy, respectively. In the half 
metal ($z < 0$), we set
$\hat H =  \sum_{\pm} (p^2/2 m_{\pm} - \varepsilon_{{\rm F},\pm})
P_{\pm}$, where $m_{\pm}$ and $\varepsilon_{{\rm F},\pm} = \hbar^2
k_{\pm}^2/2m_{\pm}$ are the effective mass and the Fermi energy for
majority ($+$) and minority ($-$) carriers in the half metal, 
and $P_{\pm} = (1/2) \vm(x) \cdot (1 \pm \vsigma)$. We take the 
limit $\varepsilon_{{\rm F},-}
\to -\infty$, so that there are no minority carriers in the half
metal. We further assume that the interface has a 
normal-state transmission probability $\tau \ll 1$, which we model
through the presence of a potential barrier $V \delta(z)$ at the
interface.

We choose a right-handed set of unit vectors $\ve_1$, $\ve_2$, and
$\ve_3$ and consider a variation of the magnetization direction $\vm$
of the form 
\begin{equation}
  \vm(x) = ( \ve_1 \cos \phi_{\rm m}
+  \ve_2 \sin \phi_{\rm m}) \sin \theta_{\rm m}(x) + 
   \ve_3\cos \theta_{\rm m}(x).
\end{equation}
We then employ a gauge transformation that rotates $\vm$ to the
$\ve_3$-direction
\begin{equation} 
  \Psi(x) \to \left( \begin{array}{cc}
  U^{\dagger}(x) & 0 \\ 0 & U^{\rm T}(x) \end{array} \right)
  \Psi(x),  
\end{equation}
with
\begin{equation}
  U(x) = e^{i \theta_{\rm m} (\vm(x) \times \ve_{3}) \cdot \sigma/2 \sin
    \theta_{\rm m}}.
\end{equation}
This gauge transformation adds a spin-dependent
gauge potential $\vA = i \hbar U^{\dagger} \vnabla U$ to 
the Hamiltonian $\hat H$ \cite{volovik1987}, but it does not affect
the singlet superconducting order parameter, $U^{\rm T} i \sigma_2
\Delta U = i \sigma_2 \Delta$. 

Since the domain wall width $l_{\rm d}$ is typically much larger than
the Fermi wavelength, we may neglect spatial variations of $\vA$. The 
wavefunction $\Psi_{\rm e}$ of a (majority)
quasiparticle in the half metal incident on the superconductor then 
reads
\begin{equation}
  \Psi_{\rm e}(\vr) = \frac{1}{\sqrt{v_{+,z}}} e^{i k_x x + i
    k_y y} 
  \left( \begin{array}{c} e^{i k_z z} 
  + r_{\rm ee} e^{-i k_z z} \\ 0 \\ r_{\rm he}
  e^{i k_z z} \\ 0 \end{array}
  \right),
  \label{eq:wavefunc}
\end{equation}
where $r_{\rm ee}$ and $r_{\rm he}$ are the amplitudes of normal
reflection and Andreev reflection, respectively. Further $k_x = k_+ \cos
\varphi \sin \theta$, $k_y = k_+ \sin \varphi \sin \theta$, and $k_z = k_+ \cos
\theta = m_+ v_{+,z}/\hbar$, where the polar angles $\varphi$ and
$\theta$ parameterize the propagation direction of the electron with
respect to the superconductor interface and the domain wall.
We
neglected the small difference of the wavenumbers of electrons and
holes if the excitation energy $\varepsilon$ is finite. 

The Andreev reflection amplitude $r_{\rm he}$ can be found by matching
$\Psi_{\rm e}$ to a linear combination of 
the four linearly independent wavefunctions in the superconductor,
\begin{equation}
  \Psi_{\alpha,\beta}(\vr) \propto
  e^{i k_x x + i k_y y + i q(\alpha,\beta) z}
  \left( \begin{array}{c} 1 \\ -i \alpha e^{i \phi_{\rm m}}\\
  i \alpha e^{-i \eta(\beta)} \\
  e^{-i \eta(\beta) - i \phi_{\rm m}} \end{array} \right),
\end{equation}
where $\alpha,\beta = \pm 1$,
$\eta(\beta) = \phi - \phi_{\rm m} + 
\beta \arccos(\varepsilon/\Delta)$ and
$q(\alpha,\beta)$ is the solution of
\begin{eqnarray}
  q^2 &=& k_{\rm S}^2 - k_x^2 - k_y^2
  + \frac{2 i m_{\rm S} \beta}{\hbar^2} \sqrt{\Delta^2 - \varepsilon^2}
  + \alpha k_x \frac{\partial \theta_{\rm m}}{\partial x} ~~~
\end{eqnarray}
with $\mbox{Im}\, q > 0$.
With the help of the boundary conditions at the
half-metal--superconductor interface $z=0$, we then calculate the
Andreev reflection amplitude $r_{\rm he}$ to lowest order in $\partial
\theta_{\rm m}/\partial x$,
\begin{equation}
  r_{\rm he}(\theta,\varphi) =
  \frac{\tau(\theta)
  k_{+} \sin \theta \cos \varphi\,
  e^{-i (\phi - \phi_{\rm m})} \Delta }{4 (k_{{\rm S}}^2 - k_{+}^2
  \sin^2 \theta)
  \sqrt{\Delta^2-\varepsilon^2}}
  \frac{\partial \theta_{\rm m}}{\partial x},
  \label{eq:rhe}
\end{equation}
where we used the Andreev approximation (which is valid for all angles
$\theta$ if $k_{\rm S}^2 \gtrsim k_{+}^2 \gg \Delta m_{\rm
  S}/\hbar^2$) and eliminated the potential barrier $V$ at the interface in
favor of the transmission coefficient
$ \tau(\theta) = {4 \hbar^2 v_{{\rm S},z} v_{+,z}}/[{4 V^2 +
  \hbar^2 (v_{{\rm S},z} + v_{+,z})^2}]$,
with $m_{\rm S} v_{{\rm S},z} = \hbar ({k_{\rm S}^2 - k_x^2 -
k_y^2})^{1/2}$. The amplitude $r_{\rm eh}$ for Andreev reflection of a
majority hole into a majority electron is $r_{\rm eh} = - r_{\rm he}^*$.
The presence of a finite triplet Andreev reflection amplitude at a domain
wall is consistent with a previous quasiclassical analysis of the
triplet proximity effect at a domain wall in the 
limit of weak exchange
fields \cite{volkov2005, endnote25}.

The order of magnitude of the Andreev reflection amplitude 
(\ref{eq:rhe}) can
be understood from the following argument: The amplitude that the
incident majority electron is initially Andreev reflected into
a hole of opposite spin is $\sim \tau(\theta) e^{-i \phi}$. 
Since the Andreev reflected hole exists 
up to a distance $\sim 1/k_{\rm S}$
away from the position of the incident majority carrier
\cite{recher2001}, 
there is a
finite overlap with majority hole states in the half metal. The 
overlap is proportional $(\partial \theta_{\rm m}/
\partial x)/k_{\rm S}$, hence the parameter dependence of 
Eq.\ (\ref{eq:rhe}). 

\begin{figure}
\epsfxsize=0.99\hsize
\epsffile{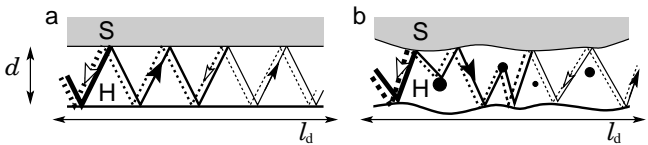}
\caption{\label{fig:3} Ballistic (a) and disordered (b) half-metallic
  film of thickness $d$ laterally coupled to a
  superconductor. The Andreev reflection amplitude in the presence of
  a slowly-varying magntization direction is enhanced by
  multiple scattering at the superconductor interface.}
\end{figure}

{\em Andreev reflection in thin half-metallic films.} We now apply the
above result to an extended half-metallic film of thickness $d$
laterally coupled to an $s$-wave superconductor, see Fig.\ \ref{fig:3}a. 
We assume that the film is in the clean limit (mean free path $\gg
l_{\rm d}$).
As before, we consider a
domain wall for which the magnetization varies in the $x$ direction,
and calculate the amplitude for Andreev reflection off this domain
wall. (There is no Andreev reflection in the absence of a domain
wall.)
The scattering states in the film are parameterized using polar angles
$\theta$ and $\varphi$ which set the magnitude of the (quantized)
momentum in the $z$ direction and the propagation direction in the
$xy$ plane, respectively. 
In the thin film geometry
electrons reflect repeatedly off the half-metal--superconductor
interface. Since the wavefunctions of the
incident electron and the Andreev reflected hole have the same
dependence on the position $\vr$, see Eq.\ (\ref{eq:wavefunc}),
amplitudes for Andreev scattering from reflections at different
positions at the interface add up coherently. This results in 
an enhancement of the Andreev reflection probability similar in origin
to the ``reflectionless tunneling effect'' in disordered
normal-metal--superconductor junctions \cite{nazarov1994b}.
Combining contributions 
from the entire width of the domain wall, we then find that the effective 
reflection amplitude for Andreev reflection off the domain wall is
\begin{equation}
  r_{\rm he}^{\rm eff}(\theta,\varphi) = 
  \frac{\tau(\theta) k_{+} \cos \theta
  e^{ - i (\phi - \phi_{\rm m})} \Delta \delta \theta_{\rm m}}{8 (k_{\rm S}^2 -
  k_{+}^2 \sin^2 \theta) d \sqrt{\Delta^2-\varepsilon^2}} 
  \mbox{sign}\,(\cos \varphi),
  \label{eq:rheeff}
\end{equation}
where $\delta \theta_{\rm m} = 
\theta_{\rm m}(\infty) - \theta_{\rm m}(-\infty)$ is the total angle 
by which the
magnetization direction changes. The same result is found by directly
solving the scattering problem in the thin-film geometry 
\cite{unpublished}. For thin films, this Andreev
reflection amplitude is significantly larger than the single
reflection amplitude of Eq.\ (\ref{eq:rhe}).

Equation (\ref{eq:rheeff}) is the main result of this letter. As
advertised in our introduction,
the effective Andreev reflection amplitude is independent of the
width $l_{\rm d}$ of the domain wall, its location, and the
angle of incidence $\varphi$. (The absence of a dependence on
$\varphi$ implies that the Andreev reflection amplitude does not
depend on the orientation of the domain wall.)
The appearance of the azimuthal angle $\phi_{\rm m}$ in
the scattering phase is consistent with the Andreev reflection
amplitude found in Ref.\ \cite{beri2009} for the 
serial geometry (see also Ref.\ \cite{braude2007}) \cite{endnote26}.

{\em Applications.} As an application, we now consider the conductance
$G_{\rm HS}$
of a lateral half-metal--superconductor junction (as in Fig.\
\ref{fig:1}b) and the Josephson effect in a lateral
superconductor--half-metal--superconductor junction. As before, we
consider the case that there is a domain wall somewhere below the
superconductor(s) for which the direction of the magnetization
variation is
parallel to the superconductor interface, and that the transmission 
coefficient of the half-metal--superconductor interface $\tau \ll 1$.
We also assume that the half metal is in the clean limit 
\cite{endnote27}
 and that $k_+ d \gg 1$. In order to simplify 
our final expressions, we set $k_{\rm S} = k_{+}$. We then find
\begin{eqnarray}
  G_{\rm HS}(\varepsilon) &=& \frac{2 e^2}{h} 
  \mbox{tr}\, r_{\rm he}^{\rm eff}(\varepsilon) 
    r_{\rm he}^{\rm eff}(\varepsilon)^{\dagger}
  \nonumber \\ &=&
  \frac{e^2 W}{h d} \frac{\langle \tau(\theta)^2 \rangle \Delta^2}
  {64 \pi (\Delta^2 - \varepsilon^2)} (\delta
  \theta_{\rm m})^2,
  \label{eq:G}
\end{eqnarray}
where the trace is taken over all transverse modes,
$W$ is the width of the half-metallic film, and
the brackets $\langle \ldots \rangle$ denote an angular average.

  
When calculating the Josephson effect, we take the junction to be
reflection symmetric, with a domain wall below each
superconductor such that the azimuthal angles $\phi_{\rm m}$ for the 
domain walls are equal and the angle changes $\delta \theta_{\rm m}$ 
are opposite. We then calculate the
zero-temperature supercurrent from the expression \cite{brouwer1997e}
\begin{eqnarray}
  I &=& - \frac{2 e}{\pi \hbar}
  \frac{\partial}{\partial \phi}
  \mbox{Re}\, \int_0^{\infty} d\omega
  \nonumber \\ && \mbox{} \times
  \mbox{tr}\, \ln [1 + e^{-2 \omega L/\hbar v}
  |r_{\rm he}^{\rm eff}(i \omega)|^2 e^{i \phi}],
\end{eqnarray}
where $v$ is the propagation velocity of a transverse mode, $L$ the
distance between the domain walls, and $\phi$ the phase difference
between the superconducting order parameters. For short junctions,
$\Delta \gg \hbar v_+/L$, we then find
\begin{equation}
  e I = \pi G_{\rm HS}(0) \Delta \sin \phi  ,
\end{equation}
where $G_{\rm HS}(0)$ is the Fermi level conductance of a single
half-metal--superconductor interface given in Eq.\ (\ref{eq:G})
above. For a long junction, $\Delta \ll \hbar v_+/L$ one has
\begin{equation}
  e I = \frac{8}{15} G_{\rm HS}(0) \frac{\hbar v_{+}}{L}
  \sin \phi, \label{eq:Ilong}
\end{equation}
where $v_{+} = \hbar k_{+}/m_{+}$. For 
superconductor--normal-metal--superconductor junctions the
Josephson current $I$ depends on the junction's normal-state 
conductance \cite{tinkham2004}.
Since the normal-state conductance is proportional to $\langle
\tau(\theta) \rangle$, not $\langle \tau(\theta)^2 \rangle$,
the difference with the half-metallic junction we consider here
is significant. We also note
that the long-junction limit of the supercurrent (\ref{eq:Ilong}) is
larger than the supercurrent in a serial geometry, which scales
proportional to $(\hbar v_{+} /L)^3 / \Delta^2$ \cite{beri2009}.
The junction becomes a ``$\pi$-junction'', with a supercurrent
proportional to $-\sin \phi$, if the two domain walls have equal
$\delta \theta_{\rm m}$ 
\cite{endnote28}
.


{\em Conclusion.}
Although the calculations presented in this letter are for ballistic
half-metal--superconductor junctions, we expect that the
enhanced tripled proximity effect in the lateral geometry also
exists in the presence of disorder, in the same way as
reflectionless tunneling exists both in clean and disordered junctions
\cite{nazarov1994b}. As long as the
non-Andreev reflected electron is transmitted through the domain wall,
as in Fig.\ \ref{fig:3}b, the coherent addition of amplitudes from 
multiple Andreev reflections is not affected by changes of the
 electron's propagation direction in a disordered half metallic film.
We have thus
identified a mechanism by which domain walls in a lateral geometry
contribute to the long-range proximity effect irrespective of their 
position,
orientation, and width.

We thank J.\ Bardarson, S.\ Basu, B.\ B\'eri, and D.\ Ralph for 
discussions. This work was supported by the Cornell Center for 
Materials Research under NSF Grant No. 0520404 and by the NSF under 
Grant No. DMR 0705476.


\begin{thebibliography}{26}
\bibitem{endnote24}{We here exclude interfaces between ferromagnets and superconductors with unconventional order parameters in which the Cooper pairs are not spin singlets.}
\bibitem{endnote27}{Normal reflection of majority carriers may occur at the ``edge'' of the lateral contacts, but this does not affect our result if the transparency $\tau $ of the half-metal--superconductor interface is small.}
\end{thebibliography}

\end{document}